\documentclass[11pt,twoside]{article}
\usepackage{adassconf,psfig}

\newcommand{\thingtreebase}[2]{{\em #1}-{#2}}
\newcommand{\thingtree}[1]{\thingtreebase{#1}{tree}}
\newcommand{\thingtrees}[1]{\thingtreebase{#1}{trees}}
\newcommand{\kdtree}{\thingtree{kd}}
\newcommand{\mrkdtree}{\thingtree{mrkd}}

\newcommand{\mrkdtrees}{\thingtrees{mrkd}}
\def\ie{{\it i.e.} }

\def\ux{\underline{x}}
\def\ph{\hat{p}}

\begin{document}   

\paperID{O3.2}

\title{Multi-Tree Methods for Statistics on Very Large Datasets in Astronomy}
%\titlemark{ }
\author{Alexander G. Gray, Andrew W. Moore}
\affil{School of Computer Science, Carnegie Mellon University}
%Pittsburgh, PA 15213}
\author{Robert C. Nichol}
\affil{Department of Physics, Carnegie Mellon University}
%Pittsburgh, PA 15213}
\author{Andrew J. Connolly}
\affil{Department of Physics and Astronomy, University of Pittsburgh}
\author{Christopher Genovese, Larry Wasserman}
\affil{Department of Statistics, Carnegie Mellon University}
%Pittsburgh, PA 15260}
\contact{Alexander G. Gray}
\email{agray@cs.cmu.edu}
\paindex{Gray, A. G.}
\aindex{Moore, A. W.} 
\aindex{Nichol, R. C.} 
\aindex{Connolly, A. J.} 
\aindex{Genovese, C.}
\aindex{Wasserman, L.}
\authormark{Gray, Moore, Nichol, Connolly, Genovese, \& Wasserman}

\begin{abstract}          
Many fundamental statistical methods have become critical tools for
scientific data analysis yet do not scale tractably to modern large
datasets.  This paper will describe very recent algorithms based on
computational geometry which have dramatically reduced the
computational complexity of 1) kernel density estimation (which also
extends to nonparametric regression, classification, and clustering),
and 2) the $n$-point correlation function for arbitrary $n$.
%These problems typify a larger class I call 'generalized $N$-body
%problems', and these new algorithms typify a new class of $N$-body
%solver which can treat many statistical problems for the first time,
%unlike existing solvers for physical $N$-body problems.
These new {\it multi-tree methods} typically yield orders of magnitude
in speedup over the previous state of the art for similar accuracy,
making millions of data points tractable on desktop workstations for
the first time.
\end{abstract}

\section{Statistics on Very Large Datasets}
Statistical inference methods are a basic component of astronomical
research.
%general, can be divided into two major types: {\it parametric}
%inference, in which the form of the data's distribution is assumed to
%be known, and {\it nonparametric} inference, which is focused upon
%making as few as assumptions as possible about the distribution of the
%data, letting the data `speak for themselves'.  The latter class is of
Nonparametric methods, in particular, make as few as assumptions as possible 
about the data's underlying distribution, and are thus of
particular relevance to scientific discovery in astronomy.
Unfortunately these tend to be much more
computationally intensive than parametric procedures.
%\ie exhibit unfavorable growth as the number of data $N$ increases.  
In the era of massive and ever-growing astronomical databases, such as
the SDSS and several others, astronomical data analysis would seem to
have already surpassed the tractable regime of nonparametric methods,
which is roughly in the tens of thousands of data points on modern
desktop workstations.  In this paper we summarize recent work in
computer science, in collaboration with astronomers and statisticians
(PiCA Group, \texttt{www.picagroup.org}) which has significantly extended
the ability of astronomers to perform nonparametric statistical
calculations with perfect or high accuracy on datasets of millions of
points and beyond.

\section{Adaptive {\kdtree} Structures}  
A {\kdtree} records a $d$-dimensional data set containing $N$ records.
Each node represents a set of data points by their bounding box.
%A node containing fewer than $\rmin$
%points is a leaf.  
Non-leaf nodes have two children, obtained by splitting the widest
dimension of the parent's bounding box.  This crucial aspect of the
construction procedure makes this data structure {\it adaptive} to the
data distribution, unlike fixed grids or other simpler tree
structures.  For the purposes of this paper, nodes are split until
they contain only one point, where they become leaves.  An {\mrkdtree}
is a conventional {\kdtree} decorated, at each node, with extra
statistics about the node's data, such as their count, centroid, and
covariance.  They are an instance of the idea of `cached sufficient
statistics' and are quite efficient in practice.  {\mrkdtrees} can be
built quickly, in time $O(dN \log d + d^2 N)$, where $d$ is the
dimensionality.

%\section{Recursive Pruning}  
Figure~\ref{fig-mrkdtree} shows an {\mrkdtree} as well as simple
examples of two mechanisms which can be used to reduce computation.
Two basic prototype problems in computational geometry are that of {\it
range-searching}, or finding all points within radius $r$ of a query
point $\ux_q$, and {\it range-counting}, in which the task is to simply
return the number of such points.  By using the bounding boxes of
subsets of the dataset associated with nodes in the tree, we can
exclude all of these subsets from further exploration, \ie recursive
traversal down the appropriate subtrees.  This is called {\it
exclusion} pruning.  In range-counting, we can additionally perform 
{\it inclusion} pruning, since we have stored the node counts as sufficient
statistics.  More complex forms of pruning are necessary for other problems.

\newcommand{\subfig}[1]{\begin{minipage}{0.20\textwidth}
\centerline{\psfig{file=#1,width=\textwidth}}
\end{minipage}}

\begin{figure}
\begin{center}
\begin{tabular}{cccc}
\subfig{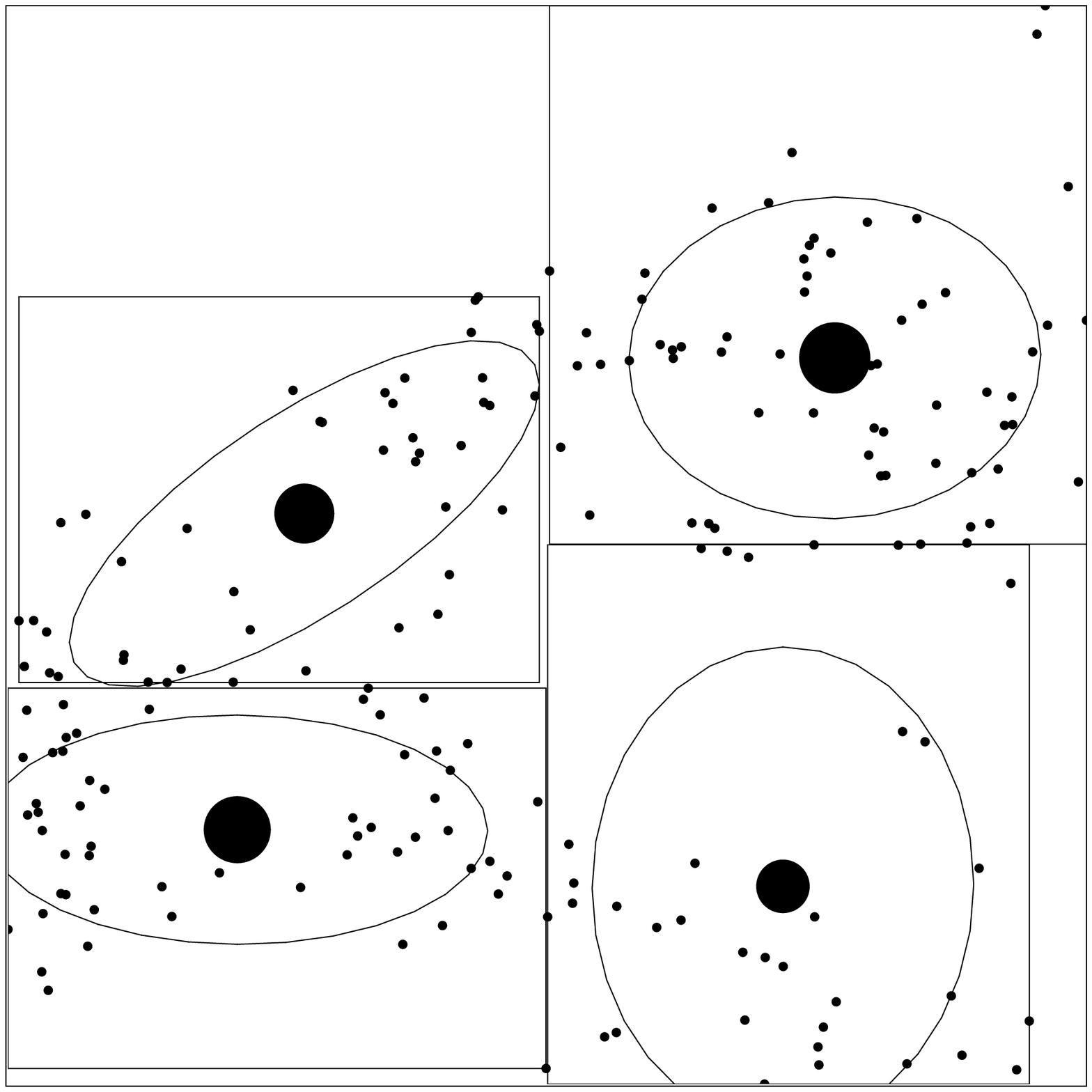} &
\subfig{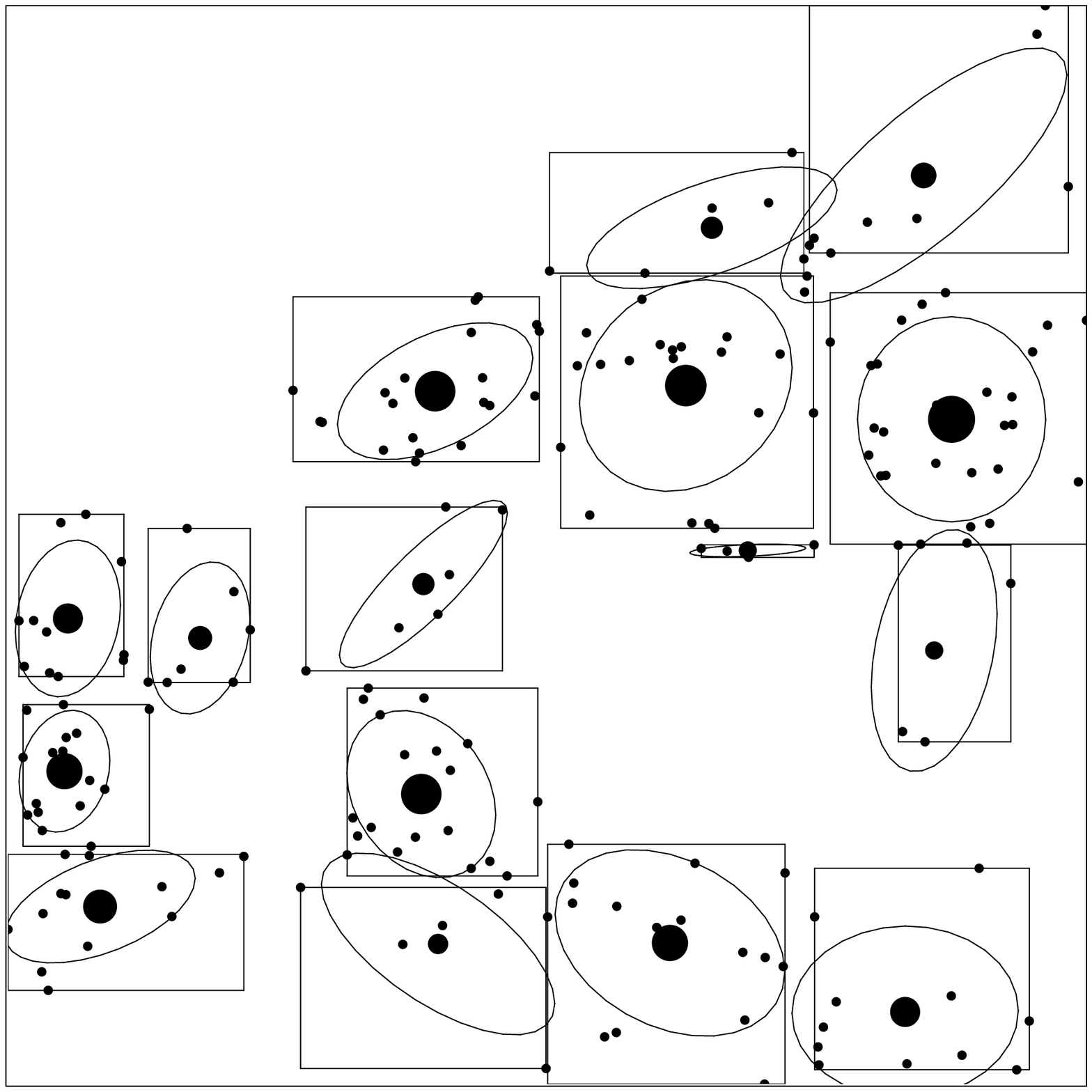} &
\subfig{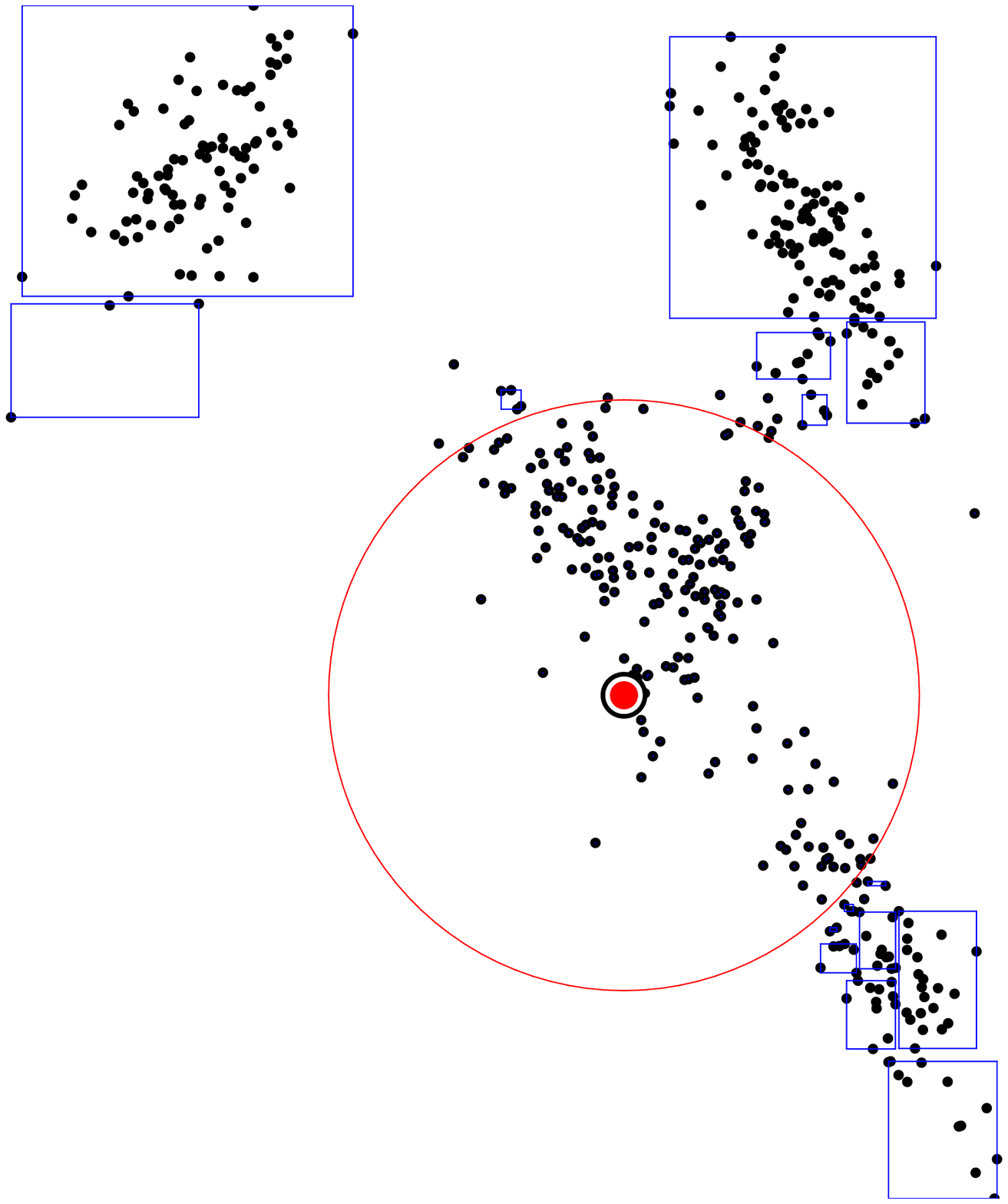} &
\subfig{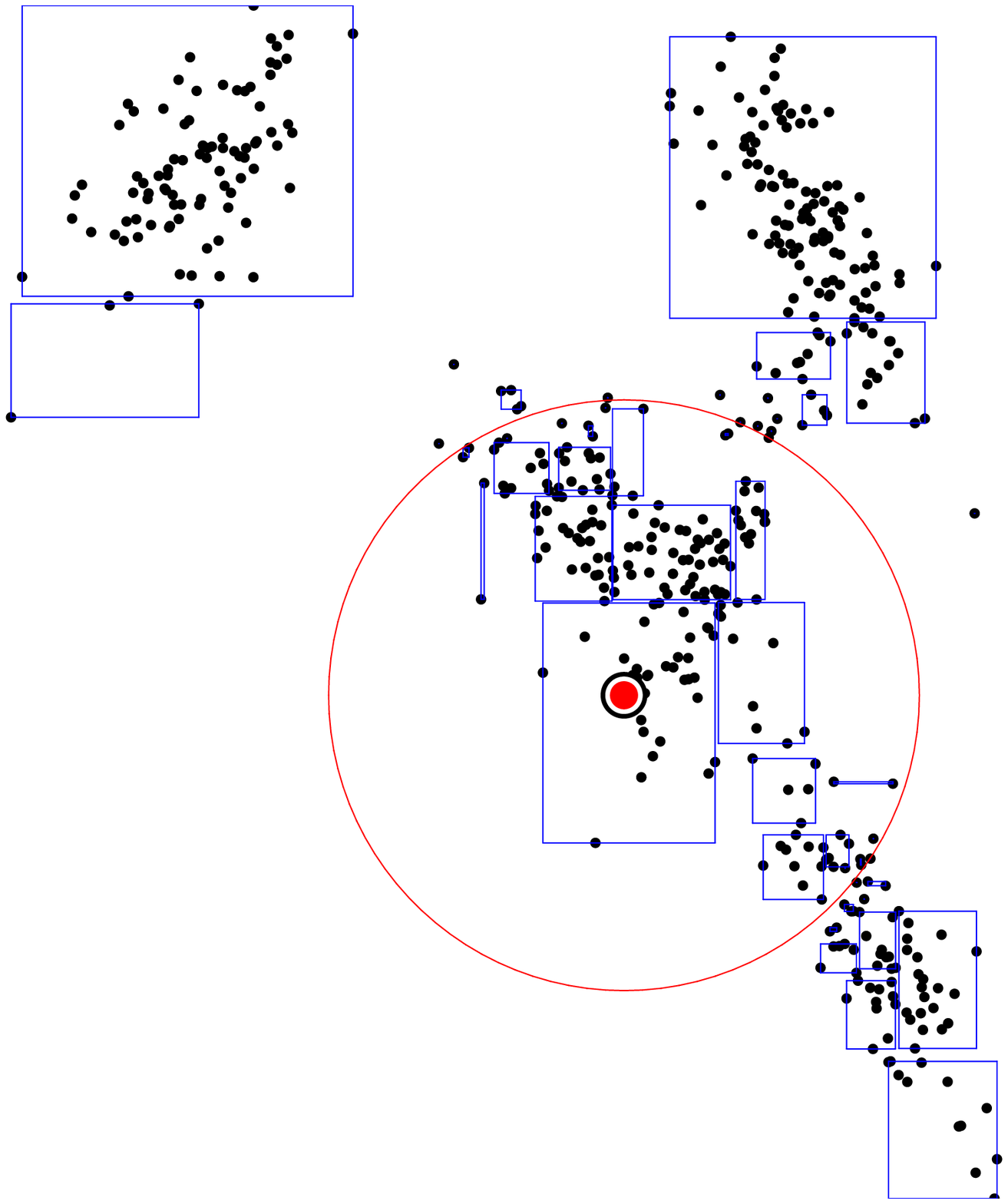}
\end{tabular}
\end{center}
\caption{\footnotesize
An {\mrkdtree}.  (a) Nodes at level 3.  (b) Nodes at level
5.  The dots are the individual data points.  The sizes and positions
of the disks show the node counts and centroids. The ellipses and
rectangles show the covariances and bounding boxes. (c) The rectangles
show the nodes pruned during a range search for one (depicted) query
and radius. (d) More pruning is possible using range-counting instead of
range-searching.\normalsize}
\label{fig-mrkdtree}
\end{figure}

\section{Multi-Tree Methods}  
Algorithms performing operations in a manner similar to that described
above have existed in computational geometry for some time.  Problems
such as computing kernel density estimates and $n$-point correlation
functions correspond to summations over pairs, triples, or in general
$n$-tuples of points.  We have developed a class of algorithms which
dramatically reduce the algorithmic complexity for such problems: it
is the extension of the previous single-tree methods to a new class we
call {\it multi-tree} methods (Gray, 2003).  The first necessary element is the
extension of the previous point-node pruning mechanisms to analogous
{\it node-node} pruning mechanisms.  This can be seen as a special
case of extending the general algorithmic device of divide-and-conquer
over a set to {\it higher-order divide-and-conquer} over
multiple sets.

\section{Kernel Density Estimation}
We first consider the method of {\it kernel density estimation} (KDE)
(Silverman 1986), a very widely analyzed and applied class of
nonparametric density estimation techniques.  Analogous kernel
estimators exist for nonparametric regression, and KDE can be used as
a subroutine to construct nonparametric classification procedures and
clustering procedures.  The task we consider in this paper is that of
computing the density estimate $\ph(\ux_q)$ for each point $\ux_q$ in
a query dataset containing $N_\mathcal{Q}$ points, given a reference
dataset containing $N_\mathcal{R}$ points and a local kernel function
$K(\cdot)$ centered upon each reference datum and having scale
parameter $h$ (the 'bandwidth'), or $K_h(\cdot)$.  The density
estimate at the $q^{\mathit th}$ query point $\ux_q$ is
\begin{equation}
\ph(\ux_q) = \frac{1}{N_\mathcal{R}} \sum_{r=1}^{N_\mathcal{R}} 
             \frac{1}{V_{dh}} K\left(\frac{\|\ux_q - \ux_r\|}{h}\right)
\end{equation}
where $d$ is the dimensionality of the data and $V_{dh} =
\int_{-\infty}^{\infty} K_h(z) dz$, 
a normalizing constant depending on $d$ and $h$.  

Note that two typical forms for the kernel function $K(\cdot)$ are the
spherical kernel ($K_h(\|\ux_q - \ux_r\|) = 1$ if $\|\ux_q - \ux_r\| <
h$, otherwise $0$, with normalizing constant $V^s_{Dh}$, the volume of
the sphere of radius $h$ in D dimensions) and the Gaussian kernel.
The spherical kernel corresponds exactly to the range-counting problem
as described earlier, but because the Gaussian function does not have
finite extent, our previous notion of pruning must be extended to one
of {\it approximation}, which will not be described here for lack of space.

\newcommand{\subfigtwo}[1]{\begin{minipage}{0.40\textwidth}
\centerline{\psfig{file=#1,width=\textwidth}}
\end{minipage}}

\begin{figure}
\begin{center}
\begin{tabular}{cc}
\subfigtwo{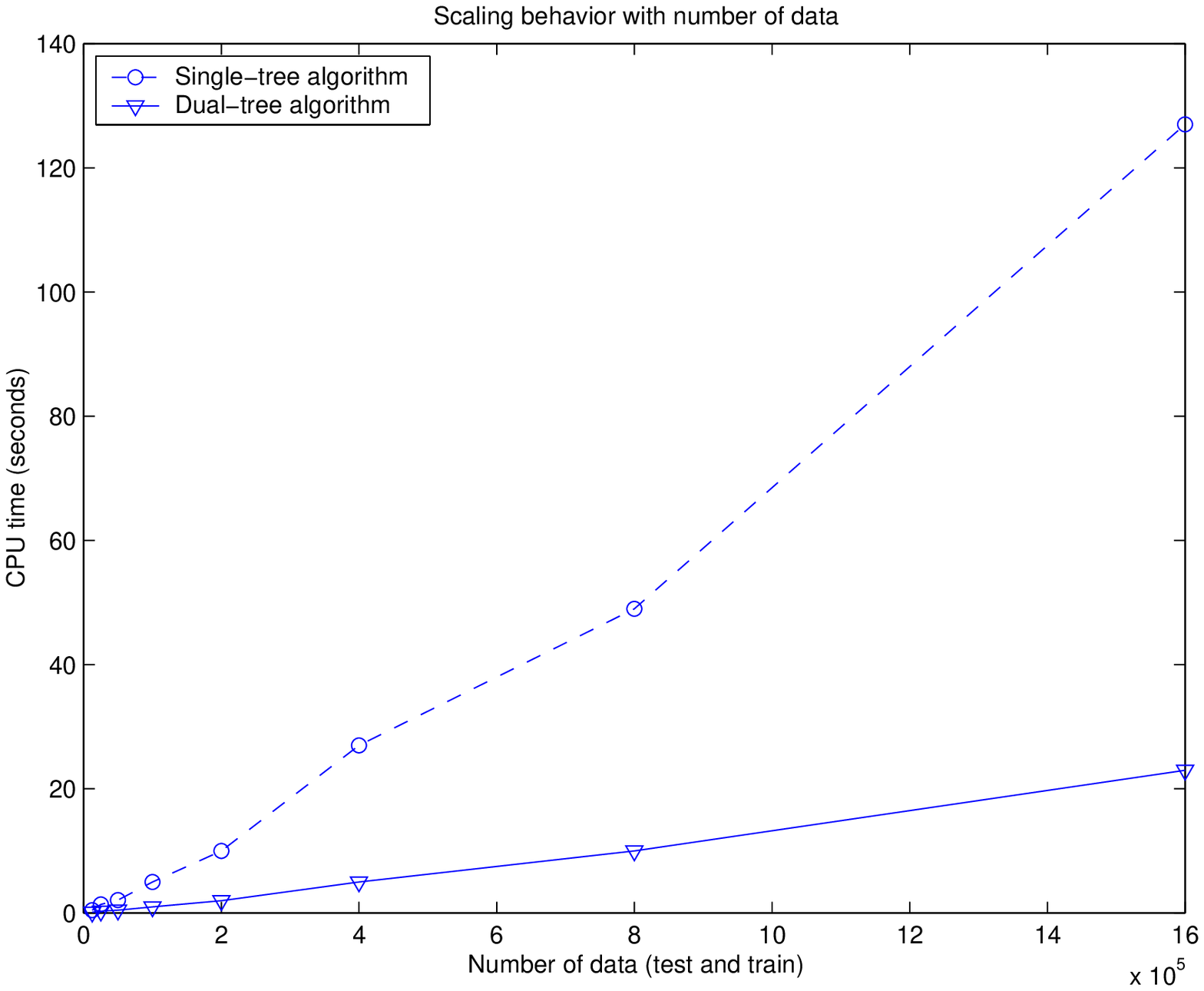} & 
\subfigtwo{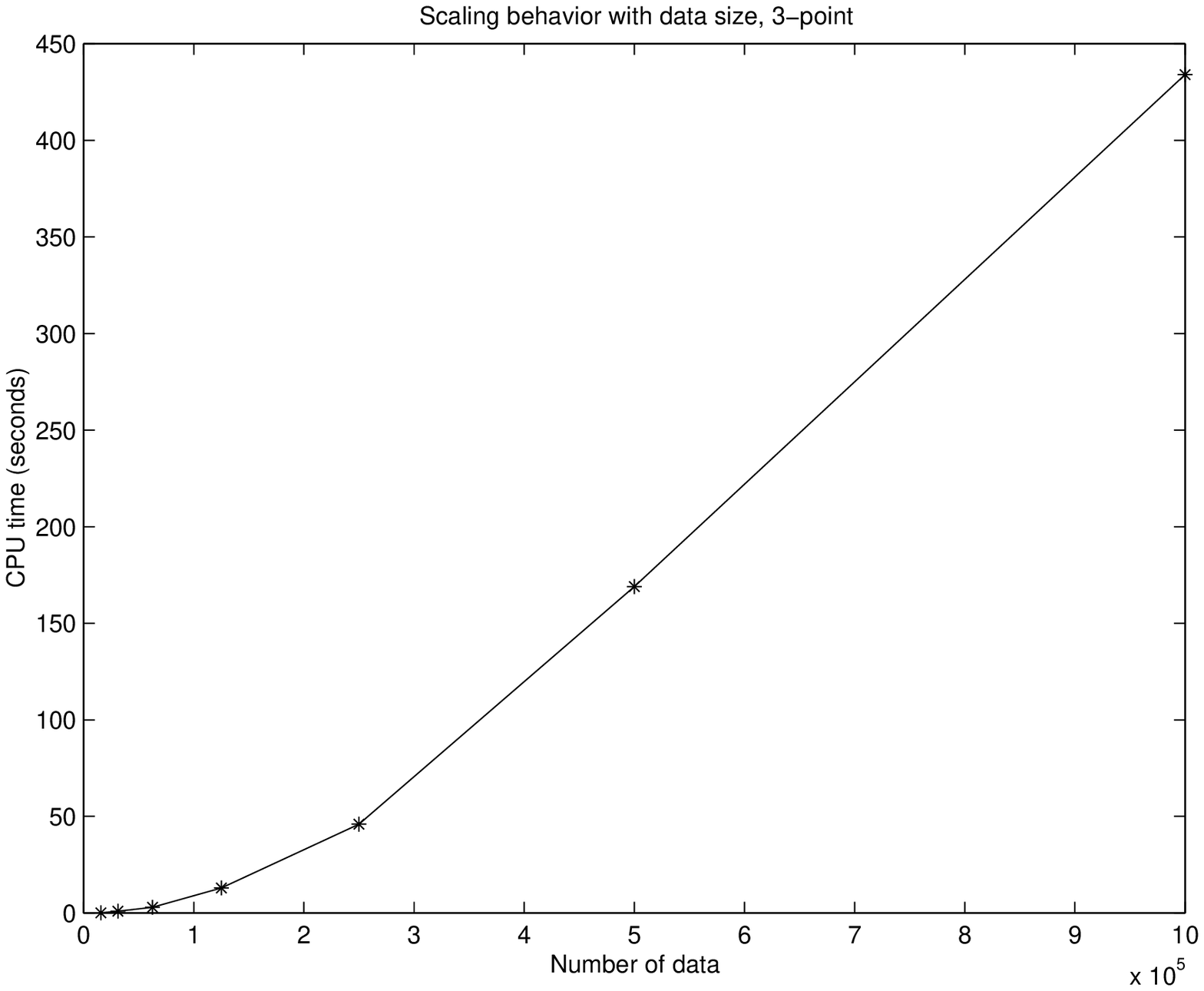}
\end{tabular}
\end{center}
\caption{\footnotesize
Examples of experimental runtimes.  (a) This shows the advantage of 
dual-tree KDE over a single-tree implementation, on an SDSS sample in 2
dimensions (RA and Dec).  Note the linear
growth in runtime for dual-tree KDE.  Performed on a 1999-era Pentium
Linux workstation. Relative approximation error is less than $10^{-6}$. 
(b) Runtime for the 3-point correlation, on a mock 
galaxy catalog based upon a Virgo Lambda CDM simulation in 3 dimensions.
Note that the computation is exact, not approximate.
Performed on a 2002-era Linux Pentium.
\normalsize}
\label{fig-perf}
\end{figure}

\section{$n$-point Correlation Functions}
{\it Point processes} are stochastic processes whose realizations
consist of point events in space (or time, the one-dimensional case).
The Poisson process is the most basic and important point process
model.  Poisson statistics thus form the foundation of spatial
statistics and have long formed a critical tool in astrophysics
(Peebles 1980).  The {\it $n$-point correlation function} ($n$pcf)
corresponds to the $n^\mathit{th}$ moment of Poisson counts.  For
example the joint probability of finding points in each of the three
volume elements $dV_q$, $dV_r$ and $dV_s$ is given by
\begin{equation}
dP = \lambda^3 dV_q dV_r dV_s [ 1 + \xi(\delta_{qr}) + \xi(\delta_{rs})
+ \xi(\delta_{sq}) + \zeta(\delta_{qr},\delta_{rs},\delta_{sq}) ]
\end{equation}
where $\delta_{qr}$, $\delta_{rs}$, and $\delta_{sq}$ are the sides of the 
triangle defined by the three points $\ux_q$, $\ux_r$, and $\ux_s$.
$\zeta()$ is called the {\it reduced} 3-point correlation function. 
In general we refer to this quantity in place of the full correlation function
since it is what we need to concern ourselves with computationally.

Computation of the $n$pcf can be viewed as a form of range-counting
problem: however here the problem is that of counting the number of
$n$-tuples whose pairwise distances match a user-specified template
for the permissible ranges.  The additional challenges posed by this
generalization from pairs (as in KDE) to $n$-tuples for arbitrary $n$
include the definition of an appropriate recursion strategy and
allowance of all possible permutations of the template $n$-gon.  These
additional complexities will not be described here for lack of space.

\section{Conclusion}

Figure \ref{fig-perf} shows some typical examples of experimental
performance, ranging up to 1 million points.  Further details,
including mathematical runtime analyses, can be found in (Gray \&
Moore 2003, Moore {\it et al.} 2001) and journal papers to appear.  We
anticipate that these algorithms will open the door to significant
astronomical analyses which could not have been suggested previously.

\end{document}